\documentclass[
]{ceurart}

\usepackage{listings}
\usepackage[utf8]{inputenc}
\usepackage{xcolor}
\usepackage{tcolorbox}
\usepackage{enumitem} 

\begin{document}

\copyrightyear{2025}
\copyrightclause{Copyright for this paper by its authors.
  Use permitted under Creative Commons License Attribution 4.0
  International (CC BY 4.0).}

\conference{Proceedings of SCOLIA 2025: The First Workshop on Scholarly Information Access (SCOLIA)}

\title{On the Alignment of Post-Publication Reviews \& Bibliometric and Altmetric Impact}
\subtitle{A Case Study on Expert Statements from the Science Media Center Germany}


\author[1, 2]{Dirk Tunger}[%
orcid=0000-0001-6383-9194,
email=dirk.tunger@th-koeln.de,
]
\fnmark[1]
\author[1]{Philipp Schaer}[%
orcid=0000-0002-8817-4632,
email=philipp.schaer@th-koeln.de,
url=https://ir.web.th-koeln.de,
]
\fnmark[1]
\address[1]{TH Köln -- University of Applied Sciences, Germany}
\address[2]{Research Center Juelich, Germany}

\fntext[1]{Both authors contributed equally.}

\begin{abstract}
In the context of academic publishing and peer review, this study investigates the relationship between post-publication expert evaluations, their agreement levels, and the subsequent scientific and public recognition of the reviewed research. Using expert statements from the Science Media Center Germany as a dataset, we analyze Research in Context reviews to examine the alignment between qualitative post-publication assessments and bibliometric as well as altmetric indicators. We employ a Large Language Model to translate unstructured expert reviews into a structured rating scheme. Furthermore, we correlate these evaluations with citation counts from the Web of Science and alternative impact metrics such as the Altmetric Attention Score, news mentions, and Mendeley readership statistics from the Altmetric Explorer. We investigate the alignment of positive or critical post-publication reviews and high or low citation or altmetric counts.

\end{abstract}


\maketitle

\section{Introduction}

Academic peer review is a fundamental element of the scientific publication and credibility system~\cite{spierHistoryPeerreviewProcess2002}. The rigorous feedback from peers who are both independent and qualified to judge the merits of scientific studies and publications advances science. It helps to guarantee a high level of quality of research. Although the current peer-review system is not without critique and alternatives like open peer reviews have been discussed~\cite{ross-hellauerWhatOpenPeer2017}, it is one of the cornerstones of modern science. 

Peer reviewers are expected to examine the rationale behind research questions and evaluate the originality, strengths, or weaknesses. Reviewers have the role of a gatekeeper, as only publications that meet a discipline's (unofficial and sometimes contradicting and arbitrary) quality standard are published. Reviewers evaluate what is relevant enough to be published.
Relevance is a central aspect of peer reviewing in this case, as topicality, research field-specific requirements, and research standards are among the most influential factors to be considered. We argue that the decisions that reviewers~\cite{jeffersonEffectsEditorialPeer2002} have to make are similar to the relevance decisions researchers have to make when searching and filtering for relevant information in their field of research, as they are multi-faceted and -layered~\cite{DBLP:conf/birws/BreuerST20,DBLP:journals/scientometrics/BreuerST22}.

However, researchers are not the only target group for scientific information. The general public and science journalists, who act as intermediaries between science and the public, are important actors in scientific communication. 
In this role, science journalists must scan the vast scientific output to find groundbreaking studies that are potentially relevant to the public, a challenge Herbert Simon described as ``a wealth of information creates a poverty of attention''~\cite{Simon:Organizations}. 
To mitigate this information crisis for journalists and the general public, specialized research information providers such as the Science Media Center (SMC) Germany
created services around scientific information.
One of these services are ``Research in Context'' (RIC) reviews, where exclusive embargo agreements with publishers worldwide are used to provide the public with additional post-publication reviews.
An inside view that is typically hidden by the blind peer review process. As the reviews are not aimed to guide a decision about the acceptance and publication of a manuscript, the reviewers are more free in their judgment, as their praise or critique does not directly impact the publication process anymore -- This decision is already over. We call them post-publication reviews to underline this special case.

While RIC and the other SMC services aim to provide early and recent access to science, the scientific system relies on credibility and recognition systems that mostly use citations but also mentions in social media and the like, known as Altmetrics. Previous research has shown how these two forms of scientific attributions go hand in hand, but very little research was conducted that connects the outcome and content of peer reviews with the later scientific and public attribution, measured by citations and altmetrics, respectively. 

With the help of an exclusive data set of RIC post-publication reviews on scientific studies, we would like to work along the following research questions:

\begin{enumerate}
    \item[RQ1] How can a Large Language Model transfer unstructured qualitative post-publication reviews of experts into a quantitative rating scheme? 
    How good is the inter-annotator agreement for different configurations of this transfer process? 
    \item[RQ2] Is there an alignment of positive or critical post-publication reviews and high or low citation rates? Do these alignments also correspond with altmetrics?
\end{enumerate}

\section{Datasets and Methods}

\subsection{Research in Context Post-publication Reviews}
The SMC Germany is a non-profit organization funded by the Klaus Tschira Trust. Its mission is to support journalists by providing access to exclusive pre-publication versions of scientific studies, so-called fact sheets on complex topics curated by specialists, or specific services around recent research and peer reviewing. 
Academic publishers like Springer, Nature, etc., have established a system that provides journalists with information about upcoming studies in advance under a press embargo (between 2-5 days). The studies are not yet publicly available, but journalists can get early access after registration and verification. In addition to access to the studies themselves, the SMC curates the ``Research in Context'' (RIC) service (now called ``Statements''\footnote{\url{https://www.sciencemediacenter.de/angebote?story_type=Statements} (last accessed: 26 March 2025)}, where experts in the field help to provide insights into cutting-edge research results and to provide journalists with orientation before reports on recent studies are written and published. Experts have first-hand access to the full texts of the studies. They must give an informed review to highlight strengths and weaknesses and provide an overall evaluation of the underlying research. This allows journalists to judge the significance of a new scientific finding by relying on the perspective of independent researchers more quickly and before the embargo period expires. After the embargo period, the RIC reviews will be published to the general public on the SMC website, where the reviewers' names are revealed. 

RIC reviews are a special case of peer reviews on scientific literature. While they are independent of classic pre-publication peer reviews, their reviewers don't have to weigh the pros and cons to come up with a clear suggestion for or against publication; their reviews are publicly visible and are not blinded. So, RIR reviewers are both more free to judge and put their finger in the wound, but they might also be more restricted due to the public nature of their reviews. However, we see these RIC reviews as an interesting research data set as we gain access to recent, real-world reviews on scientific literature without any domain-specific filtering. The articles cover various topics, such as climate research, energy, digital sciences, and medicine. The disciplines and topics addressed by the Science Media Center (SMC) are primarily those with direct social relevance and of public interest, such as medical topics and diseases like cancer, as well as environmental issues like climate change. 

\subsection{Data Acquisition and Inter-Rater Agreement}

\begin{figure}[t]
    \centering
    \includegraphics[width=\linewidth]{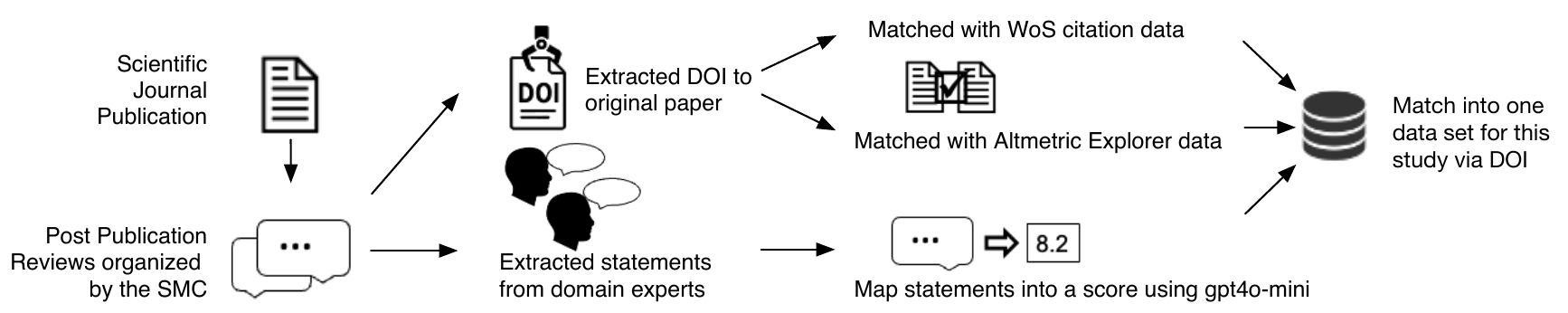}
    \caption{Data acquisition process and analysis workflow.}
    \label{fig:workflow}
\end{figure}

We crawled the website of SMC Germany in May 2024 and extracted a total of 521 RIC articles. These articles include different reviews (called statements) by domain experts on a specific primary article. We extracted the DOI or the reference to the primary article. For all 521 RIC articles, we extracted a total number of 1943 statements ($\approx3.72$ review statements per article). We ignored all editorial content around the reviewers' statements, such as highlighted quotes or subtitles, focusing only on the review statements themselves (see Figure~\ref{fig:workflow}). 

The extracted review statements are unstructured text without any score-based rating. To map the text-based reviews into a comparable score, we utilized OpenAI's \texttt{gpt4o-mini}\footnote{The code, extracted review texts, and R scripts to compute the inter-rater agreement are available here: \url{https://github.com/irgroup/SCOLIA2025-Research-in-Context} (last accessed: 26 March 2025)}. The prompt instructed the model to generate a score between 0 and 1 for each review along the following criteria:

\begin{itemize}
    \item[K0] Examination of the research question (e.g. are the aims and rationale clearly formulated?)
    \item[K1] Evaluation of originality (contribution, increase in knowledge in the literature or in the subject)
    \item[K2] The strengths and weaknesses of the method described are clearly stated
    \item[K3] Specific comments on the writing of the manuscript (e.g. spelling, organisation, illustrations, etc.)
    \item[K4] Author's interpretation of the results and conclusions drawn from the results
    \item[K5] Comments on the statistics where appropriate (e.g. whether they are robust and fit for purpose and whether the controls and sampling mechanisms are sufficiently and well described)
\end{itemize}

Values close to 0 mean that the RIC reviewers expressed weaknesses or problems in the respective evaluation criteria. Values close to 1 mean that the evaluation criteria have been fulfilled to the reviewers' satisfaction. If there is insufficient data for the evaluation criteria or a clear decision, the LLM was advised to give a value of \texttt{NA}. 
The evaluation criteria K0 to K5 were extracted from the official Elsevier guide on how to conduct a review\footnote{\url{https://www.elsevier.com/reviewer/how-to-review} (last accessed: 26 March 2025)}. 

Next, we computed the inter-rater agreement between the scores for the different reviewers per paper to measure the agreement level and identify controversial or non-controversial articles. We utilized Krippendorf's Alpha to measure the agreement as an alternative to the widely used Fleiss' Kappa values. The advantages of Alpha over Kappa statistics are that while for Kappa, all assessors have to rate the same number of subjects and use the same scale, the Alpha coefficient can usually handle more variations and computes reliabilities that are comparable across any numbers of assessors and values, different metrics, and unequal sample sizes~\cite{DBLP:conf/clef/Schaer12}. Krippendorff~\cite{https://doi.org/10.1111/j.1468-2958.2004.tb00738.x} argues for the use of Alpha in comparison to other measures because of its independence from the number of assessors and its robustness against imperfect data. No fixed or recommended values for Alpha are given, but a value of more than 0.8 is considered a (near) perfect agreement.

\subsection{Citations and Altmetrics}

With the Web of Science as developed by Eugene Garfield in the 1970s, it was possible to search for literature and see how often other scientists have cited an individual publication. Although the reasons why scientists cite each other are varied, the number of citations that a publication has received indicates how relevant other scientists consider this publication. 

Altmetrics complements traditional bibliometrics with citation statistics from social media and other online media. Thus, altmetrics can be compared to the introduction of the Science Citation Index, which enabled scientists to track where they have been cited for the first time. The only difference is that these ``citations'' are called news items, blog posts, likes, reads, shares, or Mendeley readerships: this type of altmetrics indicates how many users have saved a specific publication to their personal libraries, reflecting interest and potential future use. It ``calculates impact indicators for authors based on the number of users who stored their articles in the reference management system Mendeley''~\cite{hausteinMultidimensionalJournalEvaluation2012}. This provides early signals of scholarly engagement, often preceding and complementing traditional citation metrics.

All kinds of altmetrics make scientific impact visible more quickly than traditional bibliometrics because they evolve more quickly and dynamically.

Together, altmetrics and bibliometrics form the basis for a multidimensional view of scientific impact from different angles: Bibliometrics is the part of classical scientific communication that takes much more time to make changes visible, because every citation is a publication that has to go through the complete cycle of a paper, especially a time-consuming peer review process.

\section{Results}

\subsection{RQ1: LLM-based encoding of reviews and inter-rater agreement}

The encoding of unstructured reviews onto values between 0 and 1 for six different evaluation criteria was conducted using OpenAI's \texttt{gpt4o-mini}. We ran the experiment three times to see the range of values and to evaluate the inter-rater agreement between the different LLM runs. In the set of 521 crawled review articles, the oldest dates back to April 2016, and the newest is from March 2024. 

In a pretest based on \texttt{gpt4o} with only the first 50 RIC articles, we ran the annotation process three times in a row to compare the variance produced by the LLM. The inter-rater agreement for the three runs was 0.24, 0.2, and 0.28. 

In Table~\ref{tab:inter-rater-agreement} (left), we see the outcome of the different Alpha values on 0.2 steps for the results of all 521 RIC articles with \texttt{gpt4o-mini}. 114 articles had contradicting reviewing scores indicated by a negative Alpha value. In contrast, there were 49 articles, and all reviews were perfectly aligned. For 48 articles, we could not compute an inter-rater agreement due to missing data or parsing errors during the web crawls. The agreement averages an Alpha value of 0.27, a comparable value to the pretest that was based on the larger and more expensive \texttt{gpt4o} model\footnote{At the time of writing, the cost of \texttt{gpt4o-mini} was only 6\% of \texttt{gpt4o}.}. 

In Table~\ref{tab:inter-rater-agreement} (right), we see the results of the LLM encoding on the six criteria K0 to K5. From a total of 1943 possible single reviews, the most missing criterion is K5, with 1390. So, in 71\% of the reviews, the LLM could not find any mention of the underlying statistics. While this might be uncommon for a typical pre-publication peer review, the special setting of RIC seems to encourage reviewers to leave out these comments. Maybe because the papers under (post-)review are already accepted, and therefore, comments on these fundamental issues are already sorted out.

We can note two outcomes on RQ1: First, while there are differences in the LLM annotation outcome, it doesn't matter much with respect to the achieved inter-rater agreement in the end. Second, the cost differences between \texttt{gpt4o-mini} and \texttt{gpt4o} don't shine through on a higher inter-rater agreement. For this kind of task, cheaper and smaller models seem good enough. Nevertheless, we have to note that the general agreement between the virtual raters is not very high. Given the overall controversial nature of the underlying data, that might not be surprising, and we have to consider that we applied a rating scheme that was not mandatory for the original domain experts when writing their RIC statements. However, in a large meta-study~\cite{10.1371/journal.pone.0014331} with a total of 19,443 manuscripts, an even lower agreement with a Cohen's Kappa of 0.17 was reported, putting the numbers into perspective. 

\begin{table}[t]    
\captionabove{\emph{Left}: Inter-rater agreement for 521 RIC articles with 3.72 reviews per article. \emph{Right}: Results of the LLM-based quantification of unstructured RIC review statements. The average score is between 1~(positive reviewers' feedback) and 0~(negative feedback, serious flaws). NA indicates that there is not sufficient data to judge.}
\label{tab:inter-rater-agreement}
\begin{minipage}{.5\linewidth}
    \centering    
    \begin{tabular}{l c}
    \toprule
    Alpha & Count \\
    \midrule    
    < 0	&	114 \\
    0 < 0.2	&	90 \\
    0.2 < 0.4	&	90 \\
    0.4 < 0.6	&	91 \\
    0.6 < 0.8	&	36 \\
    0.6 < 1	&   3 \\    
    1	&   49 \\
    \midrule
    NA	&	48 \\
    \bottomrule
    \end{tabular}        
\end{minipage}%
\begin{minipage}{.5\linewidth}
    \centering
    \begin{tabular}{l c c}
    \toprule
    Criteria    &   Average & NA  \\
    \midrule
    K0 -- Research questions    &   0.76    & 467 \\
    K1 -- Originality   &   0.71    & 467 \\
    K2 -- Methods   &   0.7     & 514 \\
    K3 -- Writing   &   0.7     & 469 \\
    K4 -- Results and conclusions         &   0.71    & 500 \\
    K5 -- Statistics         &   0.57    & 1390 \\ 
    \bottomrule
    \end{tabular}    
\end{minipage}
\end{table}

\subsection{RQ2: Alignment of review scores with citations and altmetrics}

To show the impact of the present 521 publications within the scientific community and beyond in online and network media, the data retrieved from the SMC pages are enriched: Using the DOI, we searched the publication data associated with the 521 SMC peer-reviewed publications in the Web of Science and Altmetrics Explorer. We were able to do this for 405 publications. In addition to the inter-rater agreement and the mean of the six values K0 to K5, we can now provide the citations of the respective publication, the Altmetrics Attention Score, the number of news mentions 
and relate them to each other.

\begin{table}[t]
    \centering
    \caption{Bibliometric results on 521 primary studies that were discussed in RIC.}
    \begin{tabular}{l c c c c}
        \toprule
        Average K0-K5 & Papers & Paper with citations & Number of Citations & Citations per Paper  \\
        \midrule
        0 - 0.19 & 20 & 11 & 3169 & 288.1  \\
        0.2 - 0.39 & 19 & 16 & 2701 & 168.8  \\
        0.4 - 0.59 & 180 & 138 & 39,312 & 284.9  \\
        0.6 - 0.79 & 252 & 197 & 58,038 & 294.6  \\
        0.8 - 1 & 51 & 43 & 12823 & 298.2  \\
        \midrule
        total & 522 & 405 & 116,043 & 286.5  \\
        \bottomrule
    \end{tabular}
    \label{tab:Bibliometrics}
\end{table}
\begin{table}[t]
    \centering
    \caption{Altmetric results on 521 primary studies that were discussed in RIC.}
    \begin{tabular}{l c c c c}
        \toprule
        Average K0-K5 & \makecell{Altmetrics Attention\\ Score (AAS)} & \makecell{AAS\\ per Paper} & \makecell{News\\ Mentions (NM)} & \makecell{NM\\ per Paper} \\
        \midrule
        0 - 0.19 & 10,012 & 910.2 & 1,097 & 99.7\\
        0.2 - 0.39 & 16,569 & 1,035.6 & 2,227 & 139.2\\
        0.4 - 0.59 & 211,211 & 1,530.5 & 20,592 & 149.2 \\
        0.6 - 0.79 & 302,953 & 1,537.8 & 37,175 & 188.7  \\
        0.8 - 1 & 71,331 & 1,658.9 & 8,310 & 193.3 \\
        \midrule
        total & 612,076 & 1,511.3 & 69,401 & 171.4 \\
        \bottomrule
    \end{tabular}
    \label{tab:altmetrics_data}
\end{table}

The results of traditional bibliometrics in Table~\ref{tab:Bibliometrics} give a somewhat mixed picture here, with no clear trend discernible. For example, the citation rate in the two highest classes of K mean, i.e. 0.6 - 0.79 and 0.8 - 1, is higher than the average value, but even in the lowest class 0 - 0.19 the citation rate is not different from the average. Therefore, from our point of view, the picture here is indifferent.


Altmetrics are different, as we see in Table~\ref{tab:altmetrics_data}: While bibliometrics determine the perception of science within the scientific community, altmetrics focus on the attention paid to science in online media. In practice, this means that to receive a citation in bibliometrics, a completely new scientific paper that has successfully completed the long peer review process and is also published in a scientific journal is required. This process is very time-consuming~\cite{clermontDoesCitationPeriod2021}.
Altmetrics is entirely different. The perception measured here relates to news sites on the Internet and sources such as Facebook, all online sources that have not been reviewed by peers. It is, therefore, much quicker to obtain attention to a scientific publication via the sources measured with the help of Altmetrics.

If we look at the results, we can clearly see that both the Altmetrics Attention Score per paper and News Mentions per paper increase the higher the averaged K0 to K5 values get. In other words, the clearer and more positive the RIC reviews by the peers, the stronger the measured perception per paper. This is particularly evident in the two highest classes, but the overall trend across all classes is also very clear here. This confirms both the selection of papers and topics by the SMC and the assessments of the peers surveyed. It is not really surprising that this effect is measurable in the online media in particular, as this is precisely the field of application of the assessments for which the SMC's assessments are obtained.

The described effect does not apply to Mendeley Readerships (bookmarks on a scientific paper). This is because this parameter, which is one of the altmetrics, is more firmly anchored in science in terms of content and also shows a corresponding reaction here.

\section{Conclusions and Future Work}

In summary, based on this small case study, we can say that there is a correlation between the evaluation of a paper by peers, less if you look at bibliometric parameters and more if you look at altmetrics. 
We argue that the results of this case study might help to understand what different criteria might be underlying when deciding on the relevance of scientific articles. A clear limitation of our approach is that the sample of Research in Context reviews is only a proxy for these decisions and that we needed to rely on LLM encodings to analyze the unstructured reviews. Although we can see reasonable LLM-based annotations when checking random results, we could not re-assess the whole document set. Nevertheless, our proposed pipeline can describe what articles are controversial, where reviewers agree or disagree, and how these measured criteria relate to a later reception in academia and social media: Two different user groups that obviously give different kinds of recognition.

In a broader sense, we think that this work can lead to a better understanding of relevance decisions in information access, as we presented in our previous work~\cite{DBLP:journals/scientometrics/BreuerST22}. Based on the observations that       decisions encoded in information retrieval test collections overlap with high citation rates/altmetric scores, we implemented this as a beneficial service in the form of a rank fusion approach~\cite{breuerBibliometricDataFusion2023}.

In future work, we would like to investigate the underlying dynamics further. What is the specific effect of the six reviewed criteria? What of those caused the most disagreement? Is there a different effect for different scientific disciplines? How is the influence of newer LLMs, like DeepSeek-V3 or OpenAI's o3?

\section*{Acknowledgment}
The bibliometric data used in this paper are based on the local installation of Web of Science from the Competence Network for Bibliometrics located at DZHW in Berlin. The altmetric data used originate from the Altmetrics Explorer of Altmetric.com. 
We want to thank Meik Bittkowski for providing access and inside information on RIC and Nils Grote for crawling the SMC websites and conducting the initial LLM-based prestudy. 
\bibliography{bibliography}

\end{document}